\documentclass[aps,pra,superscriptaddress,nofootinbib,twocolumn]{revtex4}
\usepackage{bm}
\usepackage{graphicx,amsmath,latexsym,amssymb}
\usepackage{color}
\usepackage{dcolumn}

\begin{document}
\title{Dipole-dipole interaction in cavity-QED: perturbative regime}
\author{M. Donaire}
\email{manuel.donaire@uva.es, mad37ster@gmail.com}
\affiliation{Departamento de F\'isica Te\'orica, At\'omica y \'Optica and IMUVA,  Universidad de Valladolid, Paseo Bel\'en 7, 47011 Valladolid, Spain.}
\affiliation{On leave from Laboratoire Kastler Brossel, UPMC-Sorbonnes Universit\'es, CNRS, ENS-PSL Research University,  Coll\`{e}ge de France.}
\author{J.M. Mu\~{n}oz-Casta\~{n}eda}
\email{jose.munoz.castaneda@upm.es}
\affiliation{Departamento de F\'isica, ETSIAE,  Universidad Polit\'ecnica de Madrid, Spain}
\author{L.M. Nieto}
\email{luismiguel.nieto.calzada@uva.es}
\affiliation{Departamento de F\'isica Te\'orica, At\'omica y \'Optica and IMUVA,  Universidad de Valladolid, Paseo Bel\'en 7, 47011 Valladolid, Spain}
\begin{abstract}
We compute the interaction energies of a two-atom system placed in the middle of a perfectly reflecting planar cavity, in the perturbative regime.
Explicit expressions are provided for the van der Waals potentials of two polarisable atomic dipoles as well as for the electrostatic potential of two induced dipoles. For the van der Waals potentials,
several scenarios are considered, namely, a pair of atoms in their ground states, a pair of atoms both excited, and a pair of dissimilar atoms with one of
them excited. In addition, the corresponding phase-shift of the two-atom wavefunction is calculated in each case. The effects of the two-dimensional confinement of the electromagnetic field by the cavity are analyzed in each scenario.
\end{abstract}
\maketitle

\section{Introduction}

Modifying the interaction of atoms with the electromagnetic (EM) field by means of a cavity is at the origin of cavity-QED \cite{Bermanbook}. In the first
place, a perfectly reflecting cavity reduces the density of EM states accessible to the spontaneous emission of a single atom.
This results in an enhancement of the atomic lifetime as well as in a shift of the atomic levels. On the other hand, the strong coupling between the
cavity modes and the atomic charges drives the coherent exchange of excitations between the atom and the cavity field. Ultimately, these effects make
possible the coherent manipulation of quantum states, the entanglement between separated quantum systems \cite{expofHarocheBruneJMRaimon,ReviewHarocheetal}, and the storage of quantum information.

Considering the cavity as a macroscopic system hardly affected by the presence of the atoms inside, the interaction of the free EM field with the cavity plates can be integrated out in an effective Hamiltonian. The resultant EM interactions of the atoms are commonly referred to as cavity-assisted interactions \cite{expofHarocheBruneJMRaimon,BuhmannScheel}. This is a good approximation as long as the excitation and emission spectra of the atoms and the cavity material do not overlap, and as long as the time resolution of observation is much larger than the time of flight of photons from the atoms to the cavity plates.
Under these conditions, transient transitions average out and the effective cavity-assisted interactions become stationary.
The net result is that photon states get \emph{dressed} by multiple scattering processes with the cavity plates, and so does the photon propagator.
It is the modes of the dressed EM field which are commonly referred to as cavity modes of the cavity field.

\begin{figure}[h]
\includegraphics[height=2.7cm,width=8.7cm,clip]{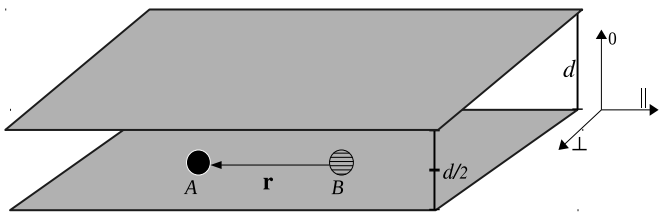}
\caption{Sketch of the setup of the problem.}\label{fig0QED}
\end{figure}

For the case of a perfectly reflecting planar cavity, both the atomic level shifts \cite{Barton,Belov,HindsSandoghdar,Jhe,Lutken} and the modified lifetimes
\cite{Milonnibook,MilonniKnight,Nha} of an excited atom have been profusely studied theoretically, and probed experimentally \cite{Haroche1}.
In this article we concentrate on the cavity-assisted dipole-dipole interactions between two atoms placed in the middle of a perfectly reflecting cavity--see Fig.\ref{fig0QED}. This is a common setup in the generation of quantum entanglement with Rydberg atoms \cite{expofHarocheBruneJMRaimon,Haroche}. In contrast to previous approaches, our calculation applies to any interatomic distance.
We compute several quantities of interest in the non-degenerate, perturbative regime. These are, the van der Waals (vdW) potentials for the case that both atoms are in
their ground states, for the case that both atoms are excited, and for the case that one atom is excited while the other, of a different kind, is in its
ground state. In addition, we calculate the electrostatic potential between two induced atomic dipoles. We give in each case the corresponding phase-shift rate
of the two-atom wavefunction.
The vdW interaction in the non-perturbative regime will be addressed in a separate publication.

Physically, the vdW and electrostatic potentials can be observed  through the forces experienced by each atom when placed inside harmonic traps, which are
proportional to the displacements of the atoms
with respect to their equilibrium positions in the absence of interaction. On the other hand, the phase-shift of the two-atom wavefunction can be observed using
atom interferometry. For instance, it is the shift that is observed in the binary interaction of
Rydberg atoms through the measurement of population probabilities \cite{ReviewHarocheetal,Rydberg}.

Concerning our approach, we apply time-dependent quantum perturbation theory, up to fourth order, in the electric dipole approximation.
The calculation is performed in the
non-degenerate, perturbative regime, where the energy difference between the intermediate and initial atomic states  is much greater than the interaction energy itself. For the case of the interaction
between excited atoms, the excitation is assumed adiabatic with respect to the detuning between the atomic species, which is a situation commonly
encountered in experiments.

The main achievements of this work are, in the first place, the computation of the Green function for the cavity field which mediate the interaction between
two atomic dipoles placed in the middle of the cavity. The effects of the confinement of the EM field to two spatial dimensions are revealed. Second, we
find out novel expressions for the resonant components of the vdW potentials of each atom when excited. The difference between each atom's potentials as
well as their discrepancy with the phase-shift of the wavefunction are exposed.

The paper is organized as follows. In Sec.\ref{lasec2} we describe the setup of the problem and compute the relevant components of the Green function of the cavity field. In Sec.\ref{lasec3} we calculate the vdW potentials and the
phase-shifts on two-atom systems in the perturbative regime.
In Sec.\ref{lasec4} we compute the electrostatic potential of two induced dipoles.  We summarize the conclusions in Sec.\ref{lasec5}.

\section{Green function of the cavity field}\label{lasec2}

We aim at computing the cavity-assisted dipole-dipole interactions between two atoms, $A$ and $B$, placed in the middle of a perfectly reflecting planar
cavity of thickness $d$, and separated by a distance $\mathbf{r}$ along an axis parallel to the cavity plates (see Fig.\ref{fig0QED}). 
These interactions are mediated by virtual photons created and annihilated at the position of each atom. Hence the relevant quantity to be computed is the Green function of the cavity field that the atomic dipoles induce at the position of each other.

To this end, we use the effective semiclassical approach outlined in the introduction. 
 The plates of the cavity are treated as passive and semiclassical objects which reflect photons with no losses. The photons mediating the interactions are
 created at the position of the atoms by the electric field operator in the interaction
 Hamiltonian, $W$. It reads, in the electric dipole approximation, $W=W_{A}+W_{B}$, with $W_{A,B}=-\mathbf{d}_{A,B}\cdot\mathbf{E}(\mathbf{R}_{A,B})$. Here
 $\mathbf{d}_{A,B}$ are the electric dipole moment operators of each atom, $\mathbf{E}$ is the electric field operator and $\mathbf{R}_{A,B}$ are
 the classical position vectors of the atomic centers of mass, with $\mathbf{r}=\mathbf{R}_{A}-\mathbf{R}_{B}$. $W$ is considered as a perturbation to the free Hamiltonian of the atoms and the EM cavity
 field, $H_{0}=H_{A}+H_{B}+H_{EM}$, with
\begin{eqnarray}
H_{A}&=&\sum_{i}\hbar\omega^{A}_{i}|A_{i}\rangle\langle A_{i}|,\quad H_{B}=\sum_{i}\hbar\omega^{B}_{i}|B_{i}\rangle\langle B_{i}|,\nonumber\\
H_{EM}&=&\sum_{\mathbf{k},\vec{\epsilon}}\hbar\omega(a^{\dagger}_{\mathbf{k},\vec{\epsilon}}a_{\mathbf{k},\vec{\epsilon}}+1/2).
\end{eqnarray}
Here, $|A_{i}\rangle$, $|B_{i}\rangle$ denote the ith states of atoms $A$, $B$, with energies $\hbar\omega^{A}_{i}$ and $\hbar\omega^{B}_{i}$, respectively.
The operators  $a^{\dagger}_{\mathbf{k},\vec{\epsilon}}$ and $a_{\mathbf{k},\vec{\epsilon}}$ are the creation and annihilation operators of photons
of frequency $\omega=ck$, momentum $\hbar\mathbf{k}$ and polarization vector $\vec{\epsilon}$ respectively, in terms of which the electric field operators in
$W_{A,B}$ read
\begin{eqnarray}
\mathbf{E}(\mathbf{R}_{A,B})&=&\sum_{\mathbf{k}}\mathbf{E}^{(-)}_{\mathbf{k}}(\mathbf{R}_{A,B})+\mathbf{E}^{(+)}_{\mathbf{k}}(\mathbf{R}_{A,B})\label{E}\\
&=&i\sum_{\mathbf{k},\vec{\epsilon}}\sqrt{\frac{\hbar ck}{2\mathcal{V}\epsilon_{0}}}
[\vec{\epsilon}a_{\mathbf{k}}e^{i\mathbf{k}\cdot\mathbf{R}_{A,B}}-\vec{\epsilon}^{\,*}a^{\dagger}_{\mathbf{k}}e^{-i\mathbf{k}\cdot\mathbf{R}_{A,B}}],\nonumber
\end{eqnarray}
with $\mathcal{V}$ being a volume of quantization. Our semiclassical approximation with regard to the interaction between the EM field and the cavity
plates consists of assuming that the virtual
photons created at the location of one of the atoms reflect off the plates any number of times before being absorbed either by itself or by the other atom.
In each reflection process the dynamical excitation of the plates is discarded, and so is the time of flight of photons between any pair of scattering processes. Under these conditions, the
intermediate photonic states and the EM vacuum can be considered as dressed by multiple reflection processes with the cavity plates \cite{WileySipe}. 
Equivalently, the dressing can be assigned to the electric field operator within the framework of macroscopic QED \cite{BuhmannScheel}. 
The net result is the effective discretization of the modes of the EM field within the cavity. Mathematically, this is achieved by  setting 
$\mathcal{V}=\mathcal{V}_{Cav}$ in Eq.(\ref{E}) or, equivalently, by imposing ideal boundary conditions in Maxwell's equations for the electric field. Those are, 
the vanish of the components of the electric field parallel to the plates and the discontinuity of its normal component at the location of the plates. 
By doing so, the resultant components of the dyadic Green's function $\mathbb{G}$ for the electric field induced at a point  ($\mathbf{r},d/2$) 
by a non-polarisable electric dipole of frequency $\omega=ck$ placed at $(\mathbf{0},d/2)$ are \cite{Miltonbook}
\begin{align}
G_{\parallel\parallel}(\mathbf{r},d/2;k)&=\int\frac{\textrm{d}^{2}q}{2(2\pi)^{2}}e^{i\mathbf{q}\cdot\mathbf{r}}\frac{q_{x}^{2}-k^{2}}{k^{2}\rho\sin{\rho d}}(1-\cos{\rho d}),\label{gxx}\\
G_{\perp\perp}(\mathbf{r},d/2;k)&=\int\frac{\textrm{d}^{2}q}{2(2\pi)^{2}}e^{i\mathbf{q}\cdot\mathbf{r}}\frac{q_{y}^{2}-k^{2}}{k^{2}\rho\sin{\rho d}}(1-\cos{\rho d}),\label{gyy}\\
G_{00}(\mathbf{r},d/2;k)&=\int\frac{\textrm{d}^{2}q}{2(2\pi)^{2}}e^{i\mathbf{q}\cdot\mathbf{r}}\frac{q^{2}}{k^{2}\rho\sin{\rho d}}(1+\cos{\rho d}),\label{gzz}
\end{align}
where  $\mathbf{q}$ is a two-dimensional reciprocal vector parallel to the plates and $\rho\equiv\sqrt{k^{2}-q^{2}}$. The quantization axis is taken
perpendicular to the plates, which is denoted by the index 0, while the indices $\parallel$ and $\perp$ refer to axis parallel to the plates which are
parallel and perpendicular to $\mathbf{r}$, respectively --cf. Fig.\ref{fig0QED}. Off-diagonal components are all null.
According to the fluctuation-dissipation theorem, the quadratic fluctuations of the electric field in the dressed vacuum, $|\tilde{0}_{\gamma}\rangle$,
read at zero temperature
\begin{equation}
\langle\tilde{0}_{\gamma}|\mathbf{E}(\mathbf{0},d/2;k)\mathbf{E}^{\dagger}(\mathbf{r},d/2;k)|\tilde{0}_{\gamma}\rangle=\frac{-\hbar k^{2}}{\pi\epsilon_{0}}
\textrm{Im}[\mathbb{G}(\mathbf{r},d/2;k)].\label{FDT}
\end{equation}

It will be found useful in the calculations to use the identity $\sin^{-1}{\rho d}=-2i(1+e^{i\rho d})\sum_{m=1}^{\infty}e^{i\rho md}$ in order to write
$\mathbb{G}$ as an infinite power series. In doing so, it is possible to ascribe a simple physical meaning to each term of the resultant series. That is,
the term of order $m$  of each component, say $G_{jj}^{(m)}\sim e^{i\:md\rho}$, with $j=0,\parallel,\perp$, accounts for the contribution  of $m$ reflections off the
plates. In the following, we will use either formulation according to its mathematical manageability. Lastly, it is also useful to write the components of
$\mathbb{G}$ in the spherical basis, with components $\{0,+,-\}$, in order to trace the polarization of the photons which mediate the corresponding atomic
transitions, $\{\pi,\sigma^{-},\sigma_{+}\}$ respectively. The change of basis yields the following relationships,
$G_{+-}=G_{-+}=(G_{\parallel\parallel}+G_{\perp\perp})/2$, $G_{++}=G_{--}=(G_{\parallel\parallel}-G_{\perp\perp})/2$.

The imaginary parts of $G_{\parallel\parallel}$, $G_{\perp\perp}$ and $G_{00}$ derive from the poles of Eqs.(\ref{gxx}), (\ref{gyy}) and (\ref{gzz}) respectively,
\begin{widetext}
\begin{align}
\textrm{Im}[G_{\parallel\parallel}(\mathbf{r},d/2;k)]&=\sum_{n=1}^{\textrm{Int}(\frac{kd}{\pi})}\frac{(-1)^{n}-1}{4dk^{2}}\Bigl[\frac{n^{2}\pi^{2}}{d^{2}}J_{0}\left(r\sqrt{k^{2}-n^{2}\pi^{2}/d^{2}}\right)+\frac{\sqrt{k^{2}-n^{2}\pi^{2}/d^{2}}}{r}J_{1}\left(r\sqrt{k^{2}-n^{2}\pi^{2}/d^{2}}\right)\Bigr],\label{imgxx}\\
\textrm{Im}[G_{\perp\perp}(\mathbf{r},d/2;k)]&=\sum_{n=1}^{\textrm{Int}(\frac{kd}{\pi})}\frac{(-1)^{n}-1}{4dk^{2}}\Bigl[k^{2}J_{0}\left(r\sqrt{k^{2}-n^{2}\pi^{2}/d^{2}}\right)-\frac{\sqrt{k^{2}-n^{2}\pi^{2}/d^{2}}}{r}J_{1}\left(r\sqrt{k^{2}-n^{2}\pi^{2}/d^{2}}\right)\Bigr],\label{imgyy}\\
\textrm{Im}[G_{00}(\mathbf{r},d/2;k)]&=\frac{-1}{4d}J_{0}(kr)-\frac{1}{2k^{2}d}\sum_{n=1}^{\textrm{Int}(\frac{kd}{2\pi})}[k^{2}-4\pi^{2}n^{2}/d^{2}]\:J_{0}\left(r\sqrt{k^{2}-4n^{2}\pi^{2}/d^{2}}\right),\label{imgzz}
\end{align}
where $J_{0}$ and  $J_{1}$ are the  Bessel functions of the first kind of orders 0 and 1 respectively. As for the real parts of $G_{jj}$, making use of the
Kramers-Kronig relationship, $k^{2}\textrm{Re}[G_{jj}(k)]=\frac{2}{\pi}\int_{0}^{\infty}dk'k'^{3}\textrm{Im}[G_{jj}(k')]/(k'^{2}-k^{2})$, we obtain
\begin{align}
\textrm{Re}[G_{\parallel\parallel}(\mathbf{r},d/2;k)]&=-\sum_{n=1}^{\textrm{Int}(\frac{kd}{\pi})}\frac{(-1)^{n}-1}{4dk^{2}}\Bigl[\frac{n^{2}\pi^{2}}{d^{2}}Y_{0}\left(r\sqrt{k^{2}-n^{2}\pi^{2}/d^{2}}\right)+\frac{\sqrt{k^{2}-n^{2}\pi^{2}/d^{2}}}{r}Y_{1}\left(r\sqrt{k^{2}-n^{2}\pi^{2}/d^{2}}\right)\Bigr]\nonumber\\
&+\sum^{\infty}_{n=\textrm{Int}(\frac{kd}{\pi})+1}\frac{(-1)^{n}-1}{2\pi dk^{2}}\Bigl[\frac{n^{2}\pi^{2}}{d^{2}}K_{0}\left(r\sqrt{n^{2}\pi^{2}/d^{2}-k^{2}}\right)+\frac{\sqrt{n^{2}\pi^{2}/d^{2}-k^{2}}}{r}K_{1}\left(r\sqrt{n^{2}\pi^{2}/d^{2}-k^{2}}\right)\Bigr],\label{regxx}\\
\textrm{Re}[G_{\perp\perp}(\mathbf{r},d/2;k)]&=-\sum_{n=1}^{\textrm{Int}(\frac{kd}{\pi})}\frac{(-1)^{n}-1}{4dk^{2}}\Bigl[k^{2}Y_{0}\left(r\sqrt{k^{2}-n^{2}\pi^{2}/d^{2}}\right)-\frac{\sqrt{k^{2}-n^{2}\pi^{2}/d^{2}}}{r}Y_{1}\left(r\sqrt{k^{2}-n^{2}\pi^{2}/d^{2}}\right)\Bigr]\nonumber\\
&+\sum^{\infty}_{n=\textrm{Int}(\frac{kd}{\pi})+1}\frac{(-1)^{n}-1}{2\pi dk^{2}}\Bigl[k^{2}K_{0}\left(r\sqrt{n^{2}\pi^{2}/d^{2}-k^{2}}\right)-\frac{\sqrt{n^{2}\pi^{2}/d^{2}-k^{2}}}{r}K_{1}\left(r\sqrt{n^{2}\pi^{2}/d^{2}-k^{2}}\right)\Bigr],\label{regyy}\\
\textrm{Re}[G_{00}(\mathbf{r},d/2;k)]&=\frac{1}{4d}Y_{0}(kr)+\frac{1}{2k^{2}d}\sum_{n=1}^{\textrm{Int}(\frac{kd}{2\pi})}[k^{2}-4\pi^{2}n^{2}/d^{2}]\:Y_{0}\left(r\sqrt{k^{2}-4n^{2}\pi^{2}/d^{2}}\right)\nonumber\\
&-\frac{1}{\pi k^{2}d}\sum^{\infty}_{n=\textrm{Int}(\frac{kd}{2\pi})+1}[k^{2}-4\pi^{2}n^{2}/d^{2}]\:K_{0}\left(r\sqrt{4n^{2}\pi^{2}/d^{2}-k^{2}}\right)
,\label{regzz}
\end{align}
where $Y_{0,1}$ and $K_{0,1}$  are the Bessel functions ($Y$) and modified Bessel functions ($K$) of the second kind, of orders 0 and 1 respectively.

Alternatively, the above expressions can be written as series in powers of the number of reflections off the plates,
\begin{align}
G_{\parallel\parallel}(\mathbf{r},d/2;k)&=\frac{e^{ikr}}{-4\pi k^{2}}\Bigl[\frac{2}{r^{3}}-\frac{2ik}{r^{2}}\Bigr]-\sum_{m=1}^{\infty}(-1)^{n}\frac{ik}{2\pi}\int_{0}^{1}\textrm{d}q\:e^{iqkmd}\Bigl[\frac{1+q^{2}}{kr\sqrt{1-q^{2}}}J_{1}(kr\sqrt{1-q^{2}})-q^{2}J_{2}(kr\sqrt{1-q^{2}})\Bigr]\nonumber\\
&-\sum_{m=1}^{\infty}(-1)^{n}\frac{k}{2\pi}\int_{0}^{\infty}\textrm{d}q\:e^{-qkmd}\Bigl[\frac{1-q^{2}}{kr\sqrt{1+q^{2}}}J_{1}(kr\sqrt{1+q^{2}})+q^{2}J_{2}(kr\sqrt{1+q^{2}})\Bigr],\label{gxxn}\\
G_{\perp\perp}(\mathbf{r},d/2;k)&=\frac{e^{ikr}}{-4\pi k^{2}}\Bigl[\frac{-1}{r^{3}}+\frac{ik}{r^{2}}+\frac{k^{2}}{r}\Bigr]-\sum_{m=1}^{\infty}(-1)^{n}\frac{ik}{2\pi}\int_{0}^{1}\textrm{d}q\:e^{iqkmd}\Bigl[\frac{1+q^{2}}{kr\sqrt{1-q^{2}}}J_{1}(kr\sqrt{1-q^{2}})-J_{2}(kr\sqrt{1-q^{2}})\Bigr]\nonumber\\
&-\sum_{m=1}^{\infty}(-1)^{n}\frac{k}{2\pi}\int_{0}^{\infty}\textrm{d}q\:e^{-qkmd}\Bigl[\frac{1-q^{2}}{kr\sqrt{1+q^{2}}}J_{1}(kr\sqrt{1+q^{2}})-J_{2}(kr\sqrt{1+q^{2}})\Bigr],\label{gyyn}\\
G_{00}(\mathbf{r},d/2;k)&=\frac{e^{ikr}}{-4\pi k^{2}}\Bigl[\frac{-1}{r^{3}}+\frac{ik}{r^{2}}+\frac{k^{2}}{r}\Bigr]-\sum_{m=1}^{\infty}(-1)^{n}\frac{ik}{2\pi}\int_{0}^{1}\textrm{d}q\:e^{iqkmd}(1-q^{2})J_{0}(kr\sqrt{1-q^{2}})\nonumber\\
&-\sum_{m=1}^{\infty}(-1)^{n}\frac{k}{2\pi}\int_{0}^{\infty}\textrm{d}q\:e^{-qkmd}(1+q^{2})J_{0}(kr\sqrt{1+q^{2}}).\label{gzzn}
\end{align}
\end{widetext}

\section{Van der Waals potentials}\label{lasec3}
\begin{figure}[h]
\includegraphics[height=4.2cm,width=8.9cm,clip]{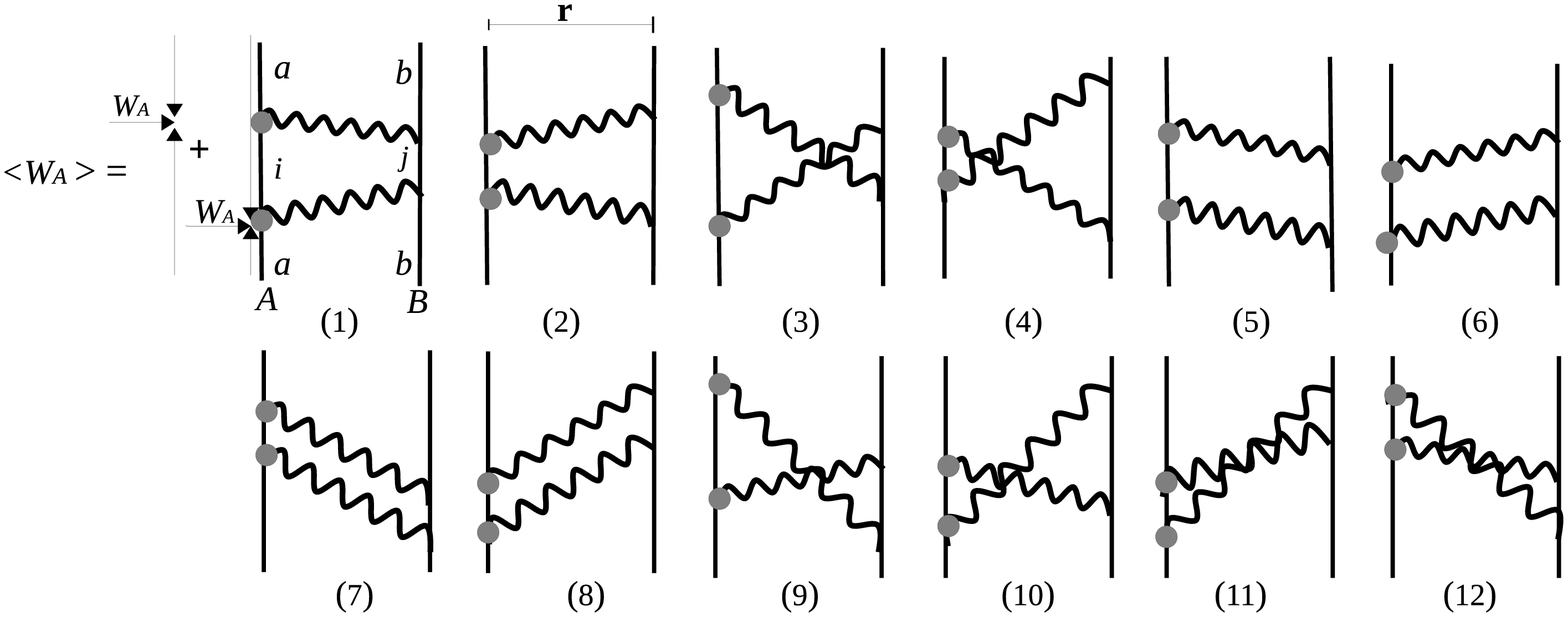}
\caption{Diagrammatic representation of the twenty-four terms which contribute to $\langle W_{A}\rangle$, two for each of the twelve diagrams.}\label{fig1QED}
\end{figure}

At leading order in time-dependent perturbation theory, twenty-four processes contribute to the vdW potentials of each atom in which a couple of photons are exchanged
between the two atoms in all possible orders, two terms for each of the diagrams in Figs.~\ref{fig1QED} and \ref{fig2QED} \cite{MePRL,MePRA,MylastPRA}.
In terms of the vdW potentials, $\langle W_{A,B}/2\rangle$, the forces on each atom are $\mathbf{F}_{A,B}=\mp\mathbf{\nabla}_{\mathbf{r}}\langle W_{A,B}/2\rangle$, respectively, with
$\mathbf{r}=\mathbf{R}_{A}-\mathbf{R}_{B}$ \cite{MePRA}. In every case, i.e., either for
ground or for excited state atoms, virtual  transitions between atomic levels are accompanied by the exchange of off-resonant photons of frequency $\omega\lesssim c/r$. Their contribution to the vdW potentials are referred to as
off-resonant vdW potentials.
In addition, for the case that one or both atoms be excited, transitions to lower energy atomic levels proceed through the exchange of photons which
resonate with the transitions. Their contribution to the vdW potentials are referred to as resonant vdW potentials \cite{WileySipe,BuhmannScheel}.
Interestingly, while the off-resonant potentials of each atom are equivalent, the resonant potentials differ \cite{MePRA,MylastPRA}.

\subsection{Off-resonant van der Waals potentials and off-resonant phase-shift}
\begin{figure}[h]
\includegraphics[height=4.2cm,width=8.9cm,clip]{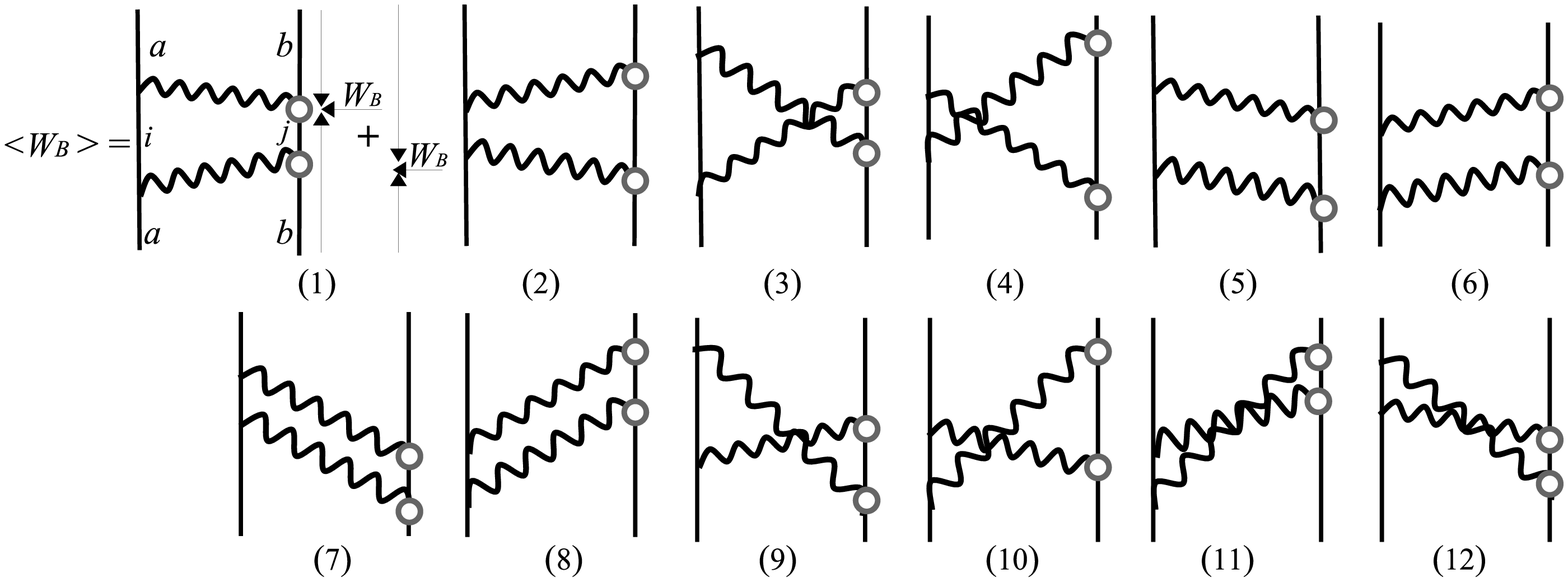}
\caption{Diagrammatic representation of the twenty-four terms which contribute to $\langle W_{B}\rangle$.}\label{fig2QED}
\end{figure}

The off-resonant component of the vdW potentials is present in the interaction between any pair of atoms regardless of whether they are in excited or
ground states.  Let us denote these states by $a$ and $b$ for atoms $A$ and $B$ respectively. The off-resonant potentials of each atom coincide,
$\langle W_{A}/2\rangle_{\textrm{off}}=\langle W_{B}/2\rangle_{\textrm{off}}$, and so does the associated phase-shift rate of the two-atom wavefunction,
$\delta\mathcal{E}_{\textrm{off}}=\langle W_{A,B}/2\rangle_{\textrm{off}}$. These energies can be also computed within the framework of stationary perturbation theory
\cite{WileySipe,McLone,Craigbook,BuhmannScheel}. It includes  both upwards and downwards virtual transitions to intermediate atomic levels.
The addition of the contributions of the twelve diagrams in Figs.~\ref{fig1QED} , \ref{fig2QED} or \ref{fig3QED}  yields  \cite{BuhmannScheel,Craigbook}
\begin{align}
\langle W_{A,B}/2\rangle_{\textrm{off}}&=\frac{-2}{\pi\hbar\epsilon_{0}^{2}c^{3}}\sum_{i,j}\int_{0}^{\infty}\textrm{d}u
\frac{u^{4}\omega_{ia}\omega_{jb}}{(u^{2}+k^{2}_{ia})(u^{2}+k^{2}_{jb})}\nonumber\\
&\times d_{ai}\cdot\mathbb{G}(\mathbf{r};iu)\cdot d_{jb}\:d_{bj}\cdot\mathbb{G}(\mathbf{r};iu)\cdot d_{ia}\nonumber\\
&=\delta\mathcal{E}_{\textrm{off}},\label{ofvdW}
\end{align}
with $\omega_{ia}=\omega^{A}_{i}-\omega^{A}_{a}$, $k_{ia}=\omega_{ia}/c$, $\omega_{jb}=\omega^{B}_{j}-\omega^{B}_{b}$, $k_{jb}=\omega_{jb}/c$,
$d_{ai}=\langle A_{a}|\mathbf{d}_{A}|A_{i}\rangle$, and $d_{bj}=\langle B_{b}|\mathbf{d}_{B}|B_{j}\rangle$.
As it stands, it suffices to substitute the expressions of the Green function components in order to calculate the off-resonant vdW potential for any
particular case. Evaluating Eqs.~(\ref{gxxn}-\ref{gzzn}) at imaginary frequencies and performing the summation over any number of reflections,
we find in the spherical basis,
\begin{align}
G_{+-}&(\mathbf{r};iu)=\frac{e^{-ur}}{8\pi u^{2}}[1/r^{3}+u/r^{2}-u^{2}/r]\nonumber\\
&+\int_{1}^{\infty}\frac{\textrm{d}\zeta}{4\pi}\frac{e^{u\zeta d}-1}{e^{2u\zeta   d}-1}u(1+\zeta^{2})J_{0}(ur\sqrt{\zeta^{2}-1}),\label{offgPMn}\\
G_{++}&(\mathbf{r};iu)=\frac{e^{-ur}}{8\pi u^{2}}[3/r^{3}+3u/r^{2}+u^{2}/r]\nonumber\\
&+\int_{1}^{\infty}\frac{\textrm{d}\zeta}{4\pi}\frac{e^{u\zeta d}-1}{e^{2u\zeta  d}-1}u(1-\zeta^{2})J_{2}(ur\sqrt{\zeta^{2}-1}),\label{offgPPn}\\
G_{00}&(\mathbf{r};iu)=\frac{e^{-ur}}{-4\pi u^{2}}[1/r^{3}+u/r^{2}+u^{2}/r]\nonumber\\
&+\int_{1}^{\infty}\frac{\textrm{d}\zeta}{2\pi}\frac{e^{u\zeta d}+1}{e^{2u\zeta d}-1}u(\zeta^{2}-1)J_{0}(ur\sqrt{\zeta^{2}-1}),\label{offgZZn}
\end{align}
where the dependence of $\mathbb{G}$ on $d/2$ has been omitted in its argument for brevity.
In all the expressions above the first terms are the components of the Green function in free-space, whereas the second terms result from multiple
scattering off the cavity plates.
As a consequence, the factor $\mathbb{G}^{2}$ in the integrand of
Eq.(\ref{ofvdW}) contains terms with two free-space factors which decay exponentially at $u\sim2/r$, terms with two multiple-scattering factors
which are exponentially suppressed at $u\sim2/d$, and terms which combine free-space and scattering factors that decay exponentially at $u\sim min(1/r,1/d)$.

The calculation of $\langle W_{A,B}/2\rangle_{\textrm{off}}$ requires the numerical integration of Eq.(\ref{ofvdW}), which depends generally on the
transition frequencies of both atoms. Nonetheless, assuming that those frequencies are roughly of the same order, say
$K\simeq k_{ia},k_{jb}$ $\forall\:i,j$, the dependence of $\langle W_{A,B}/2\rangle_{\textrm{off}}$ on the particular values of transition frequencies and dipole moments can be factored out such that approximately universal potentials can be defined as functions of $r$ and $d$ only. That is, we can write
\begin{align}
\langle W_{A,B}/2\rangle_{\textrm{off}}&\simeq\frac{-2K^{5}}{\pi\hbar\epsilon_{0}^{2}c}\sum_{i,j}\mathcal{C}_{ij}\bigl[|d_{A_{i0}}^{0}d_{B_{0j}}^{0}|^{2}V_{\textrm{off}}^{00}(r,d)\label{ofvdWeff}\\
&+(|d_{ia}^{+}d_{bj}^{+}|^{2}+|d_{ia}^{-}d_{bj}^{-}|^{2})V_{\textrm{off}}^{++}(r,d)\nonumber\\
&+(|d_{ia}^{+}d_{bj}^{-}|^{2}+|d_{ia}^{-}d_{bj}^{+}|^{2})V_{\textrm{off}}^{+-}(r,d)\bigr],\nonumber
\end{align}
where $\mathcal{C}_{ij}$ is a numerical factor of order unity whose sign is given by sgn$(\omega_{ai}\omega_{jb})$;
$d_{ia}^{p}=\langle A_{a}|d^{p}_{A}|A_{i}\rangle$ is the $p$-component of the ith transition dipole moment of atom $A$, and likewise for atom $B$; and
\begin{equation}
V_{\textrm{off}}^{pq}(r,d)=\int_{0}^{\infty}\textrm{d}q\:q^{4}G_{pq}^{2}(\mathbf{r};iKq)/[K(q^{2}+1)]^{2},\label{Vofff}
\end{equation}
with $p,q=\{+,-,0\}$, are the components in the spherical basis of the adimensional off-resonant vdW tensor potential, which depends only on $r,d$, and is independent of the internal atomic variables.

\begin{widetext}

\begin{figure}[h]
\includegraphics[height=4.cm,width=18.cm,clip]{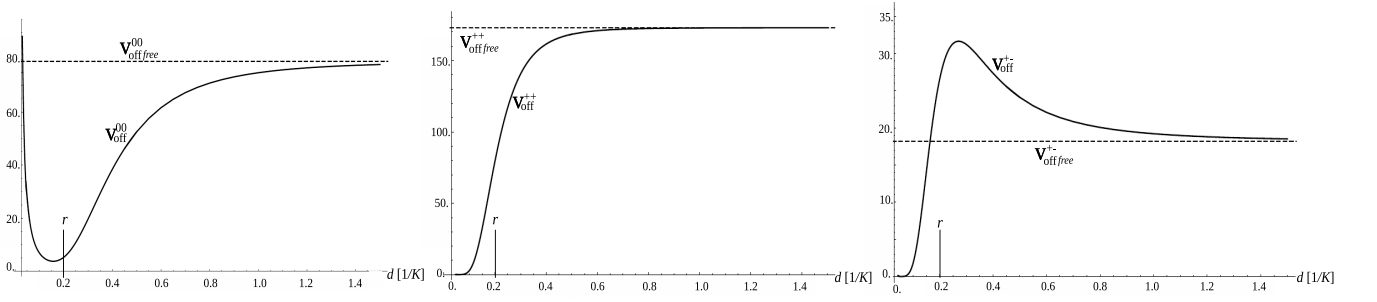}
\caption{Graphic representation of the three components of the adimensional tensor potential $\mathbb{V}_{\textrm{off}}$ of Eq.(\ref{Vofff}) as functions of $d$ for a fixed value of $r$ at $1/5K$. For comparison, the values of the analogous quantities in free space, i.e., for $d\rightarrow\infty$, are also depicted with dashed lines.}\label{figVoff}
\end{figure}

\end{widetext}

\begin{figure}[h]
\includegraphics[height=4.2cm,width=8.9cm,clip]{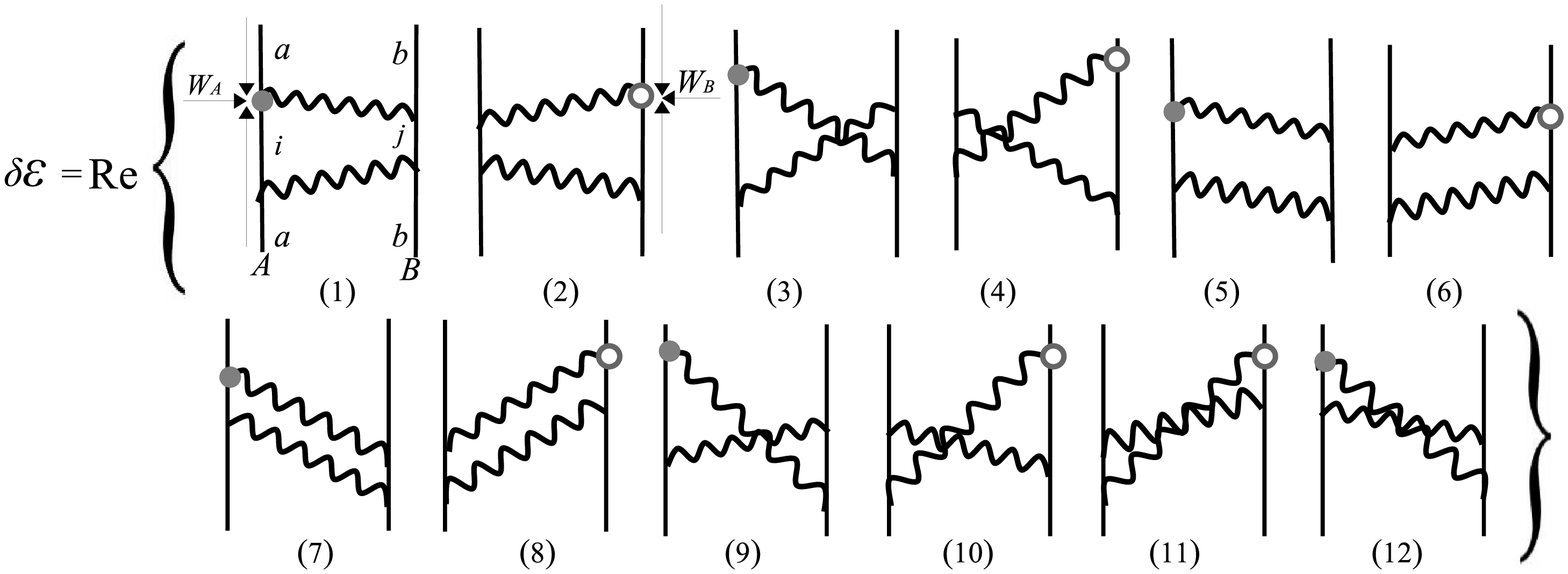}
\caption{Diagrammatic representation of the twelve terms which contribute to the phase-shift of the two-atom wavefunction, $\delta\mathcal{E}$.}\label{fig3QED}
\end{figure}

The components of $\mathbb{V}_{\textrm{off}}$ are represented in Fig.\ref{figVoff} as functions of $d$ and for a fixed value of the interatomic distance, $r=1/5K$.
Their corresponding values in free-space are also included for the sake of comparison. We observe that, as $d$ approaches $r$, the cavity confines the EM field to two effective dimensions. Interestingly, three different behaviors are found. Whereas $V_{\textrm{off}}^{++}$ decreases monotonically to zero for $d\ll r$, the component $V_{\textrm{off}}^{+-}$ shows a bump around $d\sim r$, after which it decreases to zero as well. In contrast, $V_{\textrm{off}}^{00}$ gets minimum around $d\sim r$ and increases monotonically for $d\ll r$.

\subsection{Resonant van der Waals potentials and resonant phase-shift}

In contrast to the off-resonant interaction, part of  the vdW interaction between excited atoms is mediated by virtual photons which resonate with the transitions
of one or the other atom, which is referred to as \emph{resonant interaction} \cite{WileySipe}. Correspondingly, we refer to the resonant contributions to
the potentials and to the phase-shift as resonant vdW potentials (res), and resonant phase-shift respectively. On the other hand, these resonant photons mediate also the periodic transfer of the excitation between both atoms. In the perturbative regime this transfer has a small probability proportional to $|\langle W\rangle|/\hbar|\Delta_{AB}|\ll1$, where $\Delta_{AB}$ is the detuning between the relevant transition frequencies of the atoms.
It is due to this partial and periodic transfer, as well as to the finite lifetime of excited states, that the vdW potentials with excited atoms become dynamical and
are to be computed within the framework of time-dependent perturbation theory \cite{Berman,MylastPRA,MilonniPRA,Pablo,MePRA,MePRL}. Further, for the usual case that
the excitation of the atoms be adiabatic with respect to the detuning $\Delta_{AB}$ \cite{Berman,MePRA},
the calculation simplifies to assuming that in the far past the atoms are initially excited and the interaction potential $W$ is turned on adiabatically.

\subsubsection{One atom excited}

For the case that only one of the atoms is excited, say atom $A$ at state $a>0$, while atom $B$ is in its ground state with $b=0$, only the diagrams (1) and (3) of Figs.\ref{fig1QED} and \ref{fig2QED} contribute  to the resonant potentials of atoms $A$ and $B$, respectively, yielding \cite{MePRA,MylastPRA}
\begin{align}
\langle W_{A}/2\rangle_{\textrm{res}}&=\sum_{j,i<a}\frac{2\omega_{j0}k^{4}_{ai}}{\epsilon_{0}^{2}\hbar(\omega^{2}_{ai}-\omega^{2}_{j0})}
d^{m}_{ai}d^{n}_{0j}d^{p}_{j0}d^{q}_{ia}\nonumber\\&\times\Bigl\{\textrm{Re}[G_{mn}(\mathbf{r},k_{ai})]\textrm{Re}[G_{pq}(\mathbf{r},k_{ai})]\nonumber\\
&-\textrm{Im}[G_{mn}(\mathbf{r},k_{ai})]\textrm{Im}[G_{pq}(\mathbf{r},k_{ai})]\Bigr\},\label{vdWresA}
\end{align}
\begin{align}
\langle W_{B}/2\rangle_{\textrm{res}}&=\sum_{j,i<a}\frac{2\omega_{j0}k^{4}_{ai}}{\epsilon_{0}^{2}\hbar(\omega^{2}_{ai}-\omega^{2}_{j0})}
d^{m}_{ai}d^{n}_{0j}d^{p}_{j0}d^{q}_{ia}\nonumber\\&\times\Bigl\{\textrm{Re}[G_{mn}(\mathbf{r},k_{ai})]\textrm{Re}[G_{pq}(\mathbf{r},k_{ai})]\nonumber\\
&+\textrm{Im}[G_{mn}(\mathbf{r},k_{ai})]\textrm{Im}[G_{pq}(\mathbf{r},k_{ai})]\Bigr\}.\label{vdWresB}
\end{align}
For the sake of illustration, we write in the Appendix \ref{App} the explicit expressions of the contributions of diagrams Fig.\ref{fig1QED}(3) and Fig.\ref{fig2QED}(2) to $\langle W_{A}/2\rangle_{\textrm{res}}$
and $\langle W_{B}/2\rangle_{\textrm{res}}$, respectively.

Again, assuming that those frequencies are roughly of the same order, say $K\simeq k_{ai},k_{j0}$ $\forall\:i,j$, Eqs.(\ref{vdWresA}) and (\ref{vdWresB})
can be approximated by
\begin{align}
\langle W_{A,B}/2\rangle_{\textrm{res}}&\simeq\frac{2K^{5}}{\pi\hbar\epsilon_{0}^{2}c}\sum_{i,j}\mathcal{C}^{'}_{ij}\bigl[|d_{ia}^{0}d_{0j}^{0}|^{2}
V_{A,B\textrm{res}}^{00}(r,d)\nonumber\\
&+(|d_{ia}^{+}d_{0j}^{+}|^{2}+|d_{ia}^{-}d_{0j}^{-}|^{2})V_{A,B\textrm{res}}^{++}(r,d)\nonumber\\
&+(|d_{ia}^{+}d_{0j}^{-}|^{2}+|d_{ia}^{-}d_{0j}^{+}|^{2})V_{A,B\textrm{res}}^{+-}(r,d)\bigr],\label{ofvdWresAB}
\end{align}
where  $\mathcal{C}^{'}_{ij}$ is a numerical factor of order unity, of the same sign as $\omega_{ai}-\omega_{j0}$, and the adimensional potentials read
\begin{align}
V_{A\textrm{res}}^{pq}(r,d)&=\left[\textrm{Re}^{2}[G_{pq}(\mathbf{r},K)]-\textrm{Im}^{2}[G_{pq}(\mathbf{r},K)]\right]/K^{2},\label{Vares}\\
V_{B\textrm{res}}^{pq}(r,d)&=\left[\textrm{Re}^{2}[G_{pq}(\mathbf{r},K)]+\textrm{Im}^{2}[G_{pq}(\mathbf{r},K)]\right]/K^{2},\label{Vbres}
\end{align}
with $p,q=\{+,-,0\}$. As in free-space, it is the discrepancy of the sign on the second term in each potential that gives rise to a net force on the two-atom system \cite{MePRA,Pablo,MylastPRA}.

As for the resonant phase-shift, the addition of diagrams (1) and (3) of Fig.\ref{fig3QED} is in this case
$\delta\mathcal{E}_{\textrm{res}}=\langle W_{A}/2\rangle_{\textrm{res}}$ \cite{Berman,MePRA}.

\begin{widetext}

\begin{figure}[h]
\includegraphics[height=9.cm,width=18.7cm,clip]{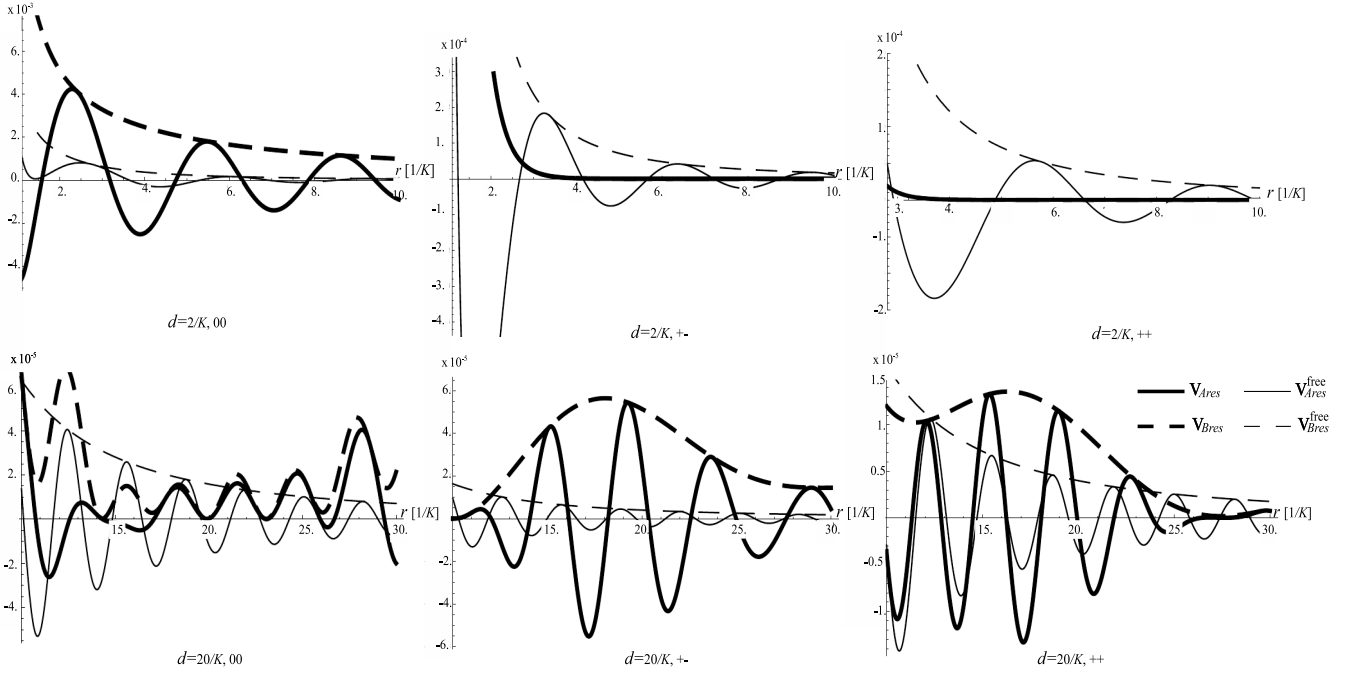}
\caption{Graphic representation of the three components of the adimensional potentials in Eqs.(\ref{Vares},\ref{Vbres}), as functions of $r$ for two different values of $d$, $2/K$ (upper inset) and $20/K$ (lower inset).}\label{Fig6Vres}
\end{figure}

\end{widetext}

The components of $\mathbb{V}_{A\textrm{res}}$ and  $\mathbb{V}_{B\textrm{res}}$ are represented in Fig.\ref{Fig6Vres}, in thick continuous and dashed
lines, respectively, as functions of $r$  for two different values of $d$, $2/K$ (upper inset) and $20/K$ (lower inset).
The free-space potentials are also depicted by thin lines for the sake of comparison. For $d\lesssim1/K$ the confinement effect of the cavity is negligible.
Generally, as already found in Refs.\cite{MePRA,Pablo}, in free space all the components of $\mathbb{V}^{\textrm{free}}_{B\textrm{res}}$ decrease
monotonically with $r$ and are the envelope of the oscillatory components of $\mathbb{V}^{\textrm{free}}_{A\textrm{res}}$ in the upper half plane.
In contrast, for $d=2/K$, the components $+-$ and $++$ of $\mathbb{V}_{A\textrm{res}}$ decrease monotonically with $r$ and coincide with those of
$\mathbb{V}_{B\textrm{res}}$. As for $d=20/K$, we observe that all the components of $\mathbb{V}_{B\textrm{res}}$ oscillate, but without changing sign.
However, whereas the pseudo-period of the oscillations of the components $++$ and $+-$ are much longer than the corresponding periods of the components
$++$ and $+-$ of  $\mathbb{V}_{A\textrm{res}}$, $V^{00}_{B\textrm{res}}$ contains faster sub-oscillations whose period is similar to the one of $V^{00}_{A\textrm{res}}$.

\subsubsection{Two dissimilar atoms excited}

When both atoms are excited, say on states $a>0$ and $b>0$,  in addition to diagrams (1) and (3), also the diagrams (2), (4), and (5-10) of
Figs.\ref{fig1QED},  \ref{fig2QED} and \ref{fig3QED} are relevant. The perturbative regime implies in this case that $|\langle W\rangle|\ll\hbar|\omega_{ai}-\omega_{jb}|$,  for any pair of intermediate states $i,j$, with $i<a,j>b$; and
$|\langle W\rangle|\ll\hbar|\omega_{bj}-\omega_{ai}|$ for any $i>a,j<b$.  For the sake of illustration, explicit expressions of the contributions of diagrams (9) and (10) of Figs.\ref{fig1QED}, \ref{fig2QED} and \ref{fig3QED} to $\langle W_{A}/2\rangle_{\textrm{res}}$, $\langle W_{B}/2\rangle_{\textrm{res}}$ and $\delta\mathcal{E}_{\textrm{res}}$, respectively, have been included in the Appendix for the case of two dissimilar atoms excited.

Generically, we can distinguish three different contributions to $\langle W_{A,B}/2\rangle_{\textrm{res}}$. These are, a first one in which the intermediate
states satisfy $i>a,j<b$, a second one in which they satisfy $i<a,j>b$, and a third one for which $i<a,j<b$. Putting them all together we have,
\begin{align}
\langle W_{A}/2\rangle_{\textrm{res}}&=\sum_{j>b,i<a}\frac{2\omega_{jb}k^{4}_{ai}}{\epsilon_{0}^{2}\hbar(\omega^{2}_{ai}-\omega^{2}_{jb})}
d^{m}_{ai}d^{n}_{bj}d^{p}_{jb}d^{q}_{ia}\nonumber\\&\times\Bigl\{\textrm{Re}[G_{mn}(\mathbf{r},k_{ai})]\textrm{Re}[G_{pq}(\mathbf{r},k_{ai})]\nonumber\\
&-\textrm{Im}[G_{mn}(\mathbf{r},k_{ai})]\textrm{Im}[G_{pq}(\mathbf{r},k_{ai})]\Bigr\}\nonumber\\
&+\sum_{j<b,i>a}\frac{2\omega_{ia}k^{4}_{bj}}{\epsilon_{0}^{2}\hbar(\omega^{2}_{bj}-\omega^{2}_{ia})}
d^{m}_{bj}d^{n}_{ai}d^{p}_{ia}d^{q}_{jb}\nonumber\\&\times\Bigl\{\textrm{Re}[G_{mn}(\mathbf{r},k_{bj})]\textrm{Re}[G_{pq}(\mathbf{r},k_{bj})]\nonumber\\
&+\textrm{Im}[G_{mn}(\mathbf{r},k_{bj})]\textrm{Im}[G_{pq}(\mathbf{r},k_{bj})]\Bigr\}\nonumber\\
&-\sum_{j<b,i<a}\frac{2\omega_{bj}k^{4}_{ai}}{\epsilon_{0}^{2}\hbar(\omega^{2}_{ai}-\omega^{2}_{bj})}
d^{m}_{ai}d^{n}_{jb}d^{p}_{bj}d^{q}_{ia}\nonumber\\&\times\Bigl\{\textrm{Re}[G_{mn}(\mathbf{r},k_{ai})]\textrm{Re}[G_{pq}(\mathbf{r},k_{ai})]\nonumber\\
&-\textrm{Im}[G_{mn}(\mathbf{r},k_{ai})]\textrm{Im}[G_{pq}(\mathbf{r},k_{ai})]\Bigr\}\nonumber\\
&+\sum_{j<b,i<a}\frac{2\omega_{ai}k^{4}_{bj}}{\epsilon_{0}^{2}\hbar(\omega^{2}_{ai}-\omega^{2}_{bj})}
d^{m}_{bj}d^{n}_{ai}d^{p}_{ia}d^{q}_{jb}\nonumber\\&\times\Bigl\{\textrm{Re}[G_{mn}(\mathbf{r},k_{bj})]\textrm{Re}[G_{pq}(\mathbf{r},k_{bj})]\nonumber\\
&+\textrm{Im}[G_{mn}(\mathbf{r},k_{bj})]\textrm{Im}[G_{pq}(\mathbf{r},k_{bj})]\Bigr\},\label{vdWresA2}
\end{align}
\begin{align}
\langle W_{B}/2\rangle_{\textrm{res}}&=\sum_{j>b,i<a}\frac{2\omega_{jb}k^{4}_{ai}}{\epsilon_{0}^{2}\hbar(\omega^{2}_{ai}-\omega^{2}_{jb})}
d^{m}_{ai}d^{n}_{bj}d^{p}_{jb}d^{q}_{ia}\nonumber\\&\times\Bigl\{\textrm{Re}[G_{mn}(\mathbf{r},k_{ai})]\textrm{Re}[G_{pq}(\mathbf{r},k_{ai})]\nonumber\\
&+\textrm{Im}[G_{mn}(\mathbf{r},k_{ai})]\textrm{Im}[G_{pq}(\mathbf{r},k_{ai})]\Bigr\}\nonumber\\
&+\sum_{j<b,i>a}\frac{2\omega_{ia}k^{4}_{bj}}{\epsilon_{0}^{2}\hbar(\omega^{2}_{bj}-\omega^{2}_{ia})}
d^{m}_{bj}d^{n}_{ai}d^{p}_{ia}d^{q}_{jb}\nonumber\\&\times\Bigl\{\textrm{Re}[G_{mn}(\mathbf{r},k_{bj})]\textrm{Re}[G_{pq}(\mathbf{r},k_{bj})]\nonumber\\
&-\textrm{Im}[G_{mn}(\mathbf{r},k_{bj})]\textrm{Im}[G_{pq}(\mathbf{r},k_{bj})]\Bigr\}\nonumber\\
&-\sum_{j<b,i<a}\frac{2\omega_{bj}k^{4}_{ai}}{\epsilon_{0}^{2}\hbar(\omega^{2}_{ai}-\omega^{2}_{bj})}
d^{m}_{ai}d^{n}_{jb}d^{p}_{bj}d^{q}_{ia}\nonumber\\&\times\Bigl\{\textrm{Re}[G_{mn}(\mathbf{r},k_{ai})]\textrm{Re}[G_{pq}(\mathbf{r},k_{ai})]\nonumber\\
&+\textrm{Im}[G_{mn}(\mathbf{r},k_{ai})]\textrm{Im}[G_{pq}(\mathbf{r},k_{ai})]\Bigr\}\nonumber\\
&+\sum_{j<b,i<a}\frac{2\omega_{ai}k^{4}_{bj}}{\epsilon_{0}^{2}\hbar(\omega^{2}_{ai}-\omega^{2}_{bj})}
d^{m}_{bj}d^{n}_{ai}d^{p}_{ia}d^{q}_{jb}\nonumber\\&\times\Bigl\{\textrm{Re}[G_{mn}(\mathbf{r},k_{bj})]\textrm{Re}[G_{pq}(\mathbf{r},k_{bj})]\nonumber\\
&-\textrm{Im}[G_{mn}(\mathbf{r},k_{bj})]\textrm{Im}[G_{pq}(\mathbf{r},k_{bj})]\Bigr\}.\label{vdWresB2}
\end{align}
Note that analogous expressions were obtained by Barcellona \emph{et al.} in Ref.\cite{Pablo} in free space.

Lastly, as for the phase shift of the two-atom wave function we find,
\begin{align}
\delta\mathcal{E}_{\textrm{res}}&=\sum_{j>b,i<a}\frac{2\omega_{jb}k^{4}_{ai}}{\epsilon_{0}^{2}\hbar(\omega^{2}_{ai}-\omega^{2}_{jb})}
d^{m}_{ai}d^{n}_{bj}d^{p}_{jb}d^{q}_{ia}\nonumber\\&\times\Bigl\{\textrm{Re}[G_{mn}(\mathbf{r},k_{ai})]\textrm{Re}[G_{pq}(\mathbf{r},k_{ai})]\nonumber\\
&-\textrm{Im}[G_{mn}(\mathbf{r},k_{ai})]\textrm{Im}[G_{pq}(\mathbf{r},k_{ai})]\Bigr\}\nonumber\\
&+\sum_{j<b,i>a}\frac{2\omega_{ia}k^{4}_{bj}}{\epsilon_{0}^{2}\hbar(\omega^{2}_{bj}-\omega^{2}_{ia})}
d^{m}_{bj}d^{n}_{ai}d^{p}_{ia}d^{q}_{jb}\nonumber\\&\times\Bigl\{\textrm{Re}[G_{mn}(\mathbf{r},k_{bj})]\textrm{Re}[G_{pq}(\mathbf{r},k_{bj})]\nonumber\\
&-\textrm{Im}[G_{mn}(\mathbf{r},k_{bj})]\textrm{Im}[G_{pq}(\mathbf{r},k_{bj})]\Bigr\}\nonumber\\
&-\sum_{j<b,i<a}\frac{2\omega_{bj}k^{4}_{ai}}{\epsilon_{0}^{2}\hbar(\omega^{2}_{ai}-\omega^{2}_{bj})}
d^{m}_{ai}d^{n}_{jb}d^{p}_{bj}d^{q}_{ia}\nonumber\\&\times\Bigl\{\textrm{Re}[G_{mn}(\mathbf{r},k_{ai})]\textrm{Re}[G_{pq}(\mathbf{r},k_{ai})]\nonumber\\
&-\textrm{Im}[G_{mn}(\mathbf{r},k_{ai})]\textrm{Im}[G_{pq}(\mathbf{r},k_{ai})]\Bigr\}\nonumber\\
&+\sum_{j<b,i<a}\frac{2\omega_{ai}k^{4}_{bj}}{\epsilon_{0}^{2}\hbar(\omega^{2}_{ai}-\omega^{2}_{bj})}
d^{m}_{bj}d^{n}_{ai}d^{p}_{ia}d^{q}_{jb}\nonumber\\&\times\Bigl\{\textrm{Re}[G_{mn}(\mathbf{r},k_{bj})]\textrm{Re}[G_{pq}(\mathbf{r},k_{bj})]\nonumber\\
&-\textrm{Im}[G_{mn}(\mathbf{r},k_{bj})]\textrm{Im}[G_{pq}(\mathbf{r},k_{bj})]\Bigr\}.\label{deltaEres2}
\end{align}

\subsubsection{Two identical atoms excited}

We consider next the case in which the two atoms are identical, $A=B$, and find in the same excited state $a>0$. The non-degenerate  condition necessary for the calculation to be perturbative reads
in this case
$|\langle W\rangle|\ll\hbar|\omega_{ai}-\omega_{ja}|$, for any pair of intermediate states $i,j$, with $i<a,j>a$.

In comparison to the case of dissimilar atoms, the only difference in the calculation is the presence of double poles when $i=j$ in the frequency integrals
which derive from the  diagrams (3), (4), (9) and (10). Explicit expressions of the contributions of diagram (4) in Figs.\ref{fig1QED} and \ref{fig3QED} to
$\langle W_{A}/2\rangle_{\textrm{res}}$ and $\delta\mathcal{E}_{\textrm{res}}$, respectively, have been included in the Appendix for the case of two identical atoms excited. As for the vdW potential, it reads,
\begin{align}
\langle W_{A}/2\rangle_{\textrm{res}}&=\sum_{i<a,j\neq i}\frac{4\omega_{ja}k^{4}_{ai}}{\epsilon_{0}^{2}\hbar(\omega^{2}_{ai}-\omega^{2}_{ja})}
d^{m}_{ai}d^{n}_{ja}d^{p}_{aj}d^{q}_{ia}\\&\times\textrm{Re}[G_{mn}(\mathbf{r},k_{ai})]\textrm{Re}[G_{pq}(\mathbf{r},k_{ai})]\nonumber\\
&+\sum_{i<a,j=i}\frac{k^{2}_{ai}}{\epsilon_{0}^{2}c\hbar} d^{m}_{ai}d^{n}_{ia}d^{p}_{ai}d^{q}_{ia}\nonumber\\
&\times\Bigl\{k_{ai}\textrm{Re}[G_{mn}(\mathbf{r},k_{ai})]\textrm{Re}[G_{pq}(\mathbf{r},k_{ai})]\nonumber\\
&-2\textrm{Re}[G_{mn}(\mathbf{r},k_{ai})]\frac{\partial}{\partial k}\Bigr[k^{2}\textrm{Re}[G_{pq}(\mathbf{r},k)]\Bigl]_{k=k_{ai}}\Bigr\},\nonumber
\end{align}
 whereas the phase shift of the two-atom wavefunction is
\begin{align}
\delta\mathcal{E}_{\textrm{res}}&=\sum_{i<a,j\neq i}\frac{4\omega_{ja}k^{4}_{ai}}{\epsilon_{0}^{2}\hbar(\omega^{2}_{ai}-\omega^{2}_{ja})}
d^{m}_{ai}d^{n}_{ja}d^{p}_{aj}d^{q}_{ia}\label{deltaEres3}\\&\times\Bigl\{\textrm{Re}[G_{mn}(\mathbf{r},k_{ai})]\textrm{Re}[G_{pq}(\mathbf{r},k_{ai})]\nonumber\\
&-\textrm{Im}[G_{ij}(\mathbf{r},k_{ai})]\textrm{Im}[G_{pq}(\mathbf{r},k_{ai})]\Bigr\}\nonumber\\
&+\sum_{i<a,j=i}\frac{k^{2}_{ai}}{\epsilon_{0}^{2}c\hbar} d^{m}_{ai}d^{n}_{ia}d^{p}_{ai}d^{q}_{ia}\nonumber\\
&\times\Bigl\{k_{ai}\textrm{Re}[G_{mn}(\mathbf{r},k_{ai})]\textrm{Re}[G_{pq}(\mathbf{r},k_{ai})]\nonumber\\
&-k_{ai}\textrm{Im}[G_{mn}(\mathbf{r},k_{ai})]\textrm{Im}[G_{pq}(\mathbf{r},k_{ai})]\nonumber\\
&-2\textrm{Re}[G_{mn}(\mathbf{r},k_{ai})]\frac{\partial}{\partial k}\Bigr[k^{2}\textrm{Re}[G_{pq}(\mathbf{r},k)]\Bigl]_{k=k_{ai}}\Bigr\}.\nonumber
\end{align}

\section{Electrostatic potential between induced dipoles}\label{lasec4}
\begin{figure}[h]
\includegraphics[height=4.2cm,width=8.9cm,clip]{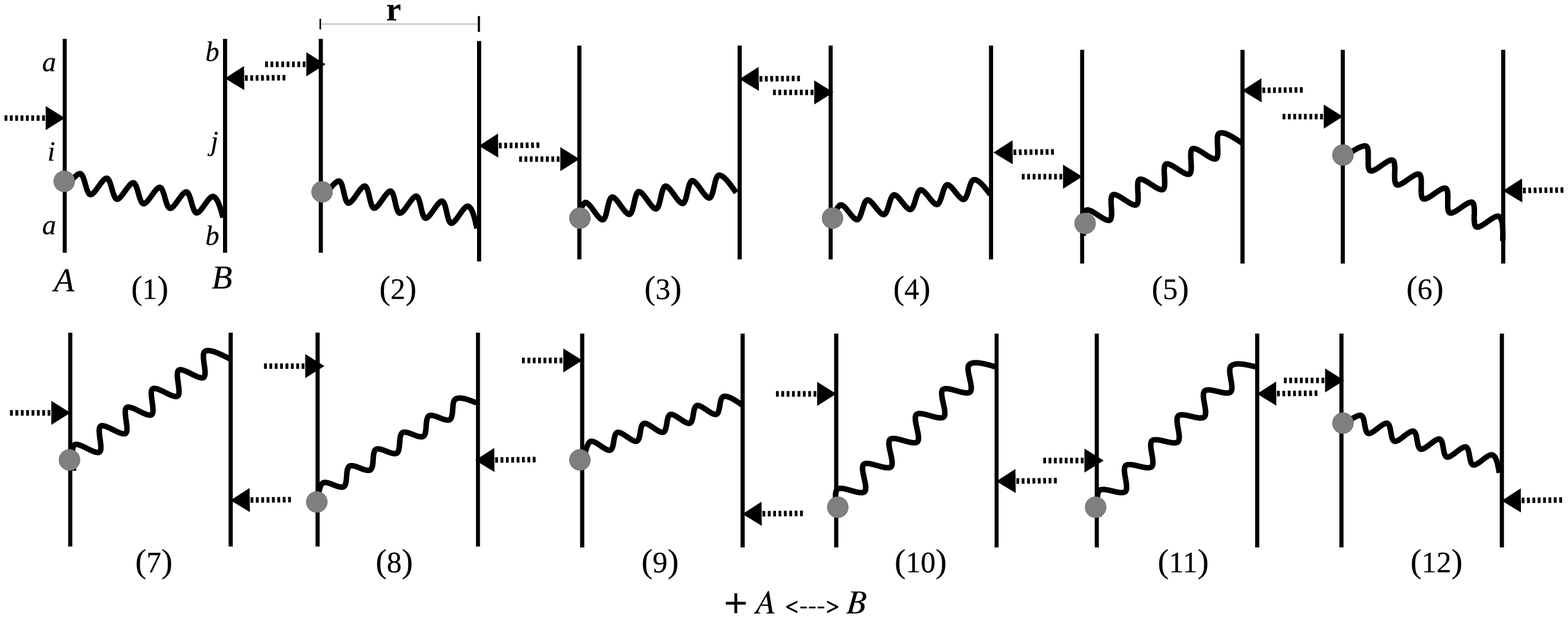}
\caption{Diagrammatic representation of the twenty-four terms (two times twelve after the exchange $A\leftrightarrow B$) which contribute to $\langle W_{A}\rangle=V_{AB}^{st}$ under the action of a constant and
uniform external field $\mathbf{E}_{0}$, whose interaction Hamiltonian $-(\mathbf{d}_{A}+\mathbf{d}_{B})\cdot\mathbf{E}_{0}$ is depicted by dashed arrows.}\label{fig4QED}
\end{figure}

\begin{figure}[h]
\includegraphics[height=4.2cm,width=8.9cm,clip]{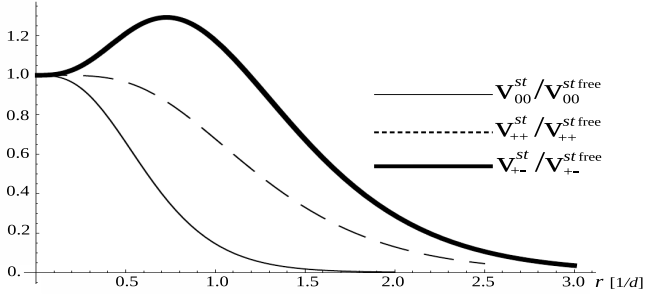}
\caption{Graphic representation of the three adimensional potentials $V^{st}_{00}$, $V^{st}_{++}$ and $V^{st}_{+-}$, as functions of $r$ and normalized by the corresponding potentials in free space.}\label{Fig7Vst}
\end{figure}

Another case of interest commonly encountered in experiments is that of the electrostatic interaction between two atomic dipoles induced by an external
static field $\mathbf{E}_{0}$. The contribution of the twenty-four diagrams of Fig.\ref{fig4QED} reduces to the electrostatic interaction between two induced
electric dipoles, $A$ and $B$, with moments $\alpha^{a}_{A}(0)\mathbf{E}_{0}$ and $\alpha^{b}_{B}(0)\mathbf{E}_{0}$, where $\alpha^{a,b}_{A,B}(0)$ are the
static polarisabilities of each atom in the states $a$ and $b$ respectively. The interaction potentials of each atom coincide in this case, and so does the
associated phase-shift rate of the two-atom wavefunction.  We denote that potential by $V^{st}_{AB}$, so that it holds
$\langle W_{A}\rangle=\langle W_{B}\rangle=\delta\mathcal{E}^{st}\equiv V^{st}_{AB}(r)$, being the forces on each atom
$\mathbf{F}_{A,B}=\mp\mathbf{\nabla}_{\mathbf{r}}V^{st}_{AB}(r)$, respectively\footnote{Note the absence of the factor 1/2 in the expression for the electrostatic
potential in comparison to the vdW potentials. This is due to the fact that the calculation is order one in $W_{A,B}$ for $V_{AB}^{st}$, while it is order two for the vdW potentials.}. The addition of all the contributions of the diagrams of Fig.\ref{fig4QED} (which refer to $ \langle W_{A}\rangle$,
in particular), yields
\begin{align}
V_{AB}^{st}(r)&=\frac{8}{\pi\epsilon_{0}c^{2}\hbar^{2}}\sum_{i,j}\frac{\langle a|\mathbf{d}|i\rangle\cdot\mathbf{E}_{0}\langle b|\mathbf{d}|j\rangle\cdot\mathbf{E}_{0}}{\omega_{ia}\omega_{jb}}\label{dd}\\
&\times\int_{0}^{\infty}\textrm{d}\omega\:\omega\textrm{Tr}\{\langle i|\mathbf{d}|a\rangle\cdot\textrm{Im}[\mathbb{G}(\mathbf{r},d/2;\omega)]\cdot\langle j|\mathbf{d}|b\rangle\}.\nonumber
\end{align}
Next, in application of the Kramers-Kronig relations on the Green function and writing its tensor components in the spherical basis, the above equation
can be written as
\begin{widetext}
\begin{align}
V_{AB}^{st}(r)&=\frac{4}{\epsilon_{0}c^{2}\hbar^{2}}\sum_{i,j}\frac{\langle a|\mathbf{d}|i\rangle\cdot\mathbf{E}_{0}\langle b|\mathbf{d}|j\rangle\cdot\mathbf{E}_{0}}{\omega_{ia}\omega_{jb}}\lim_{\omega\rightarrow0}\omega^{2}\textrm{Tr}\{\langle i|\mathbf{d}|a\rangle\cdot\textrm{Re}[\mathbb{G}(\mathbf{r},d/2;\omega)]\cdot\langle j|\mathbf{d}|b\rangle\},\nonumber\\
&=\frac{4\pi}{\epsilon_{0}\hbar^{2}d^{3}}\sum_{i,j}\frac{1}{\omega_{ia}\omega_{jb}}\Bigl[\bigl(|\langle i|d_{+}^{A}|a\rangle|^{2}|\langle j|d^{B}_{+}|b\rangle|^{2}(E^{+}_{0})^{2}+|\langle i|d_{-}^{A}|a\rangle|^{2}|\langle j|d^{B}_{-}|b\rangle|^{2}(E^{-}_{0})^{2}\bigr)V^{st}_{++}(r)\nonumber\\
&+|\langle i|d_{0}^{A}|a\rangle|^{2}|\langle j|d^{B}_{0}|b\rangle|^{2}(E^{0}_{0})^{2}V^{st}_{00}(r)+\bigl(|\langle i|d_{+}^{A}|a\rangle|^{2}|\langle j|d^{B}_{-}|b\rangle|^{2}+|\langle i|d_{-}^{A}|a\rangle|^{2}|\langle j|d^{B}_{+}|b\rangle|^{2}\bigr)E^{-}_{0}E^{+}_{0}V^{st}_{+-}(r)\Bigr],\nonumber
\end{align}
\end{widetext}
where the adimensional potentials read
 \begin{align}
 V^{st}_{00}(r)&=4\sum_{n=1}n^{2}K_{0}(\frac{2\pi r}{d}n),\nonumber\\
 V^{st}_{++}(r)&=\sum_{n=1}\frac{(-1)^{n}-1}{4}\Bigl[n^{2}K_{0}(\frac{\pi r}{d}n)+\frac{2d}{\pi r}n\:K_{1}(\frac{\pi r}{d}n)\Bigr],\nonumber\\
 V^{st}_{+-}(r)&=\sum_{n=1}\frac{(-1)^{n}-1}{4}\:n^{2}K_{0}(\frac{\pi r}{d}n),
 \end{align}
 with $K_{0}$ and $K_{1}$ being  modified Bessel functions of the second kind, of zero and first orders respectively. In the limit $r/d\rightarrow0$ the above functions converge to the free space values, $V^{st\:\textrm{free}}_{00}=d^{3}/4\pi^{2}r^{3}$, $V^{st\:\textrm{free}}_{++}=-3d^{3}/8\pi^{2}r^{3}$ and
 $V^{st\:\textrm{free}}_{+-}=-d^{3}/8\pi^{2}r^{3}$, respectively. In Fig.\ref{Fig7Vst}, the ratios between the three components of the cavity
 electrostatic potential and their corresponding values in free-space are represented as functions of $r/d$. We observe that they all go to zero as
 $r/d\gg1$. However, while the components $00$ and $++$ decrease monotonically with $r/d$, the component $+-$ presents a maximum around $r\approx d$ which
 overtakes the value in free-space.

\section{Conclusions}\label{lasec5}
In this article we have computed the dyadic Green's function of the cavity field which mediates the interaction between two atomic dipoles placed in the 
middle of a perfectly reflecting planar cavity. The components of the Green tensor are given in Eqs.(\ref{imgxx}-\ref{regzz}) and, as series in the number 
of reflection, in Eqs.(\ref{gxxn}-\ref{gzzn}).

The van der Waals potentials of each atom as well as the associated phase-shift of their wavefunction have been calculated, in the perturbative regime, for several cases of interest. These are, for the case that both atoms are in
their ground states, for the case that both atoms are excited, and for the case that one atom is excited while the other, of a different kind, is in its ground state. The discrepancies
between the resonant components of the vdW potentials and phase-shifts have been exposed for each case. In addition, we have calculated the electrostatic
potential between two induced atomic dipoles. In all the cases, the two-dimensional confinement of the EM field by the cavity arises as the cavity width
approaches the interatomic distance, and its effects depend on the polarization. A qualitative analysis of these effect has been carried out in each case, comparing the polarization-dependent potentials of the atoms within the cavity with their corresponding values in free-space [Figs.\ref{figVoff},\ref{Fig6Vres} and \ref{Fig7Vst}].

\acknowledgments
We thank M.-P. Gorza, M. Brune, J.-M. Raimond, A. Lambrecht and S. Reynaud for useful discussions on this problem.
Financial support from the French contracts ANR-10-IDEX-0001-02-PSL and ANR-13-BS04--0003-02, and from the Spanish grants MTM2014-57129-C2-1-P (MINECO) 
and VA057U16 (Junta de Castilla y Le\'on) is gratefully acknowledged.

\appendix*

\begin{widetext}

\section{Resonant contributions of some diagrams to the van der Waals potentials and phase-shifts}\label{App}
In the following, we write the expressions of the resonant contributions of some diagrams  to the vdW potentials and phase-shifts at order four in $W$. The rules to read off each diagram are as follows. The four vertices yield a tensor factor
$\frac{2\alpha_{f} c^{3}}{\pi\epsilon_{0}e^{2}}d_{ai}^{r}d_{jb}^{s}d_{jb}^{p}d_{ai}^{q}$; each wavy line contributes with a cavity field Green's function,
$k^{2}\textrm{Im}G_{pq}(\mathbf{r},k)$; and time-propagators are inserted between any pair of consecutive vertices, with time evolving from above and
from below in the far past, towards the observable vertex at instant $T$ -- as sketched in the diagrams (1) and (2) of Figs.\ref{fig1QED}, \ref{fig2QED} and \ref{fig3QED}.

For the case of one atom excited, we give the expressions of the contributions of diagrams Fig.\ref{fig1QED}(3) and Fig.\ref{fig2QED}(2) to
$\langle W_{A}/2\rangle_{\textrm{res}}$ and $\langle W_{B}/2\rangle_{\textrm{res}}$, respectively, in the adiabatic approximation. As for diagram Fig.\ref{fig1QED}(3), its contribution
to $\langle W_{A}/2\rangle_{\textrm{res}}$ is
\begin{eqnarray}
&&\frac{2\alpha_{f} c^{3}}{\pi\epsilon_{0}e^{2}}\sum_{i<a,j}d_{ai}^{r}d_{j0}^{s}d_{j0}^{p}d_{ai}^{q}\int_{-\infty}^{+\infty}
\textrm{d}k\:k^{2}\textrm{Im}G _{rs}(\mathbf{r},k)\int_{-\infty}^{+\infty}\textrm{d}k'\:k'^{2}\textrm{Im}G _{pq}(\mathbf{r},k')\nonumber\\
&\times&\int_{-\infty}^{T}\textrm{d}t\int_{-\infty}^{t}\textrm{d}t'\int_{-\infty}^{t'}\textrm{d}t''\:e^{\eta(t+t'+t'')}
\Bigl[\left(i\:e^{i\omega_{ai}T}e^{-i(T-t)\omega}e^{-i(t-t')(\omega+\omega'+\omega_{j0})}e^{-i(t'-t'')\omega'}e^{-it''\omega_{ai}}\right)+(\omega\leftrightarrow\omega')^{*}\Bigr]\nonumber\\
&=&\frac{-4\alpha_{f} c^{3}}{\pi\epsilon_{0}e^{2}}\textrm{Re}\sum_{i<a,j}d_{ai}^{r}d_{j0}^{s}d_{j0}^{p}d_{ai}^{q}\int_{-\infty}^{+\infty}
\textrm{d}k\int_{-\infty}^{+\infty}\frac{\textrm{d}k'\:k^{2}\textrm{Im}G _{rs}(\mathbf{r},k)\:k'^{2}\textrm{Im}G _{pq}(\mathbf{r},k')}
{[\omega+\omega'-(\omega_{ai}-\omega_{j0})](\omega-\omega_{ai}-i\eta)(\omega'-\omega_{ai}-i\eta)},\:\:\eta\rightarrow0^{+}.
\end{eqnarray}
The contribution of diagram Fig.\ref{fig2QED}(2) to $\langle W_{B}/2\rangle_{\textrm{res}}$ reads
\begin{eqnarray}
&&\textrm{Re}\frac{2\alpha_{f} c^{3}}{\pi\epsilon_{0}e^{2}}\sum_{i<a,j}d_{ai}^{r}d_{j0}^{s}d_{j0}^{p}d_{ai}^{q}\int_{-\infty}^{+\infty}
\textrm{d}k\:k^{2}\textrm{Im}G _{rs}(\mathbf{r},k)\int_{-\infty}^{+\infty}\textrm{d}k'\:k'^{2}\textrm{Im}G _{pq}(\mathbf{r},k')\nonumber\\
&\times&\int_{-\infty}^{T}\textrm{d}t\int_{-\infty}^{t}\textrm{d}t'\int_{-\infty}^{T}\textrm{d}t''\:e^{\eta(t+t'+t'')}
\left[\left(-ie^{i(T-t)\omega_{B}}e^{i(t-t')\omega}e^{i\omega_{A}t'}
\:e^{-i(T-t'')\omega'}e^{-it''\omega_{A}}\right)+(\omega\leftrightarrow\omega')\right]\nonumber\\
&=&\frac{4\alpha_{f} c^{3}}{\pi\epsilon_{0}e^{2}}\textrm{Re}\sum_{i<a,j}d_{ai}^{r}d_{j0}^{s}d_{j0}^{p}d_{ai}^{q}\int_{-\infty}^{+\infty}
\textrm{d}k\int_{-\infty}^{+\infty}\textrm{d}k'\frac{k^{2}\textrm{Im}G _{rs}(\mathbf{r},k)\:k'^{2}\textrm{Im}G _{pq}(\mathbf{r},k')}
{(\omega_{ai}-\omega_{j0})(\omega-\omega_{ai}-i\eta)(\omega'-\omega_{ai}+i\eta)},\:\:\eta\rightarrow0^{+}.
\end{eqnarray}

As for the case of two dissimilar atoms excited, we give the expressions of the contributions of diagrams (9) and (10) of Figs.\ref{fig1QED}, \ref{fig2QED} and \ref{fig3QED} to $\langle W_{A}/2\rangle_{\textrm{res}}$, $\langle W_{B}/2\rangle_{\textrm{res}}$ and $\delta\mathcal{E}_{\textrm{res}}$, respectively, in the adiabatic approximation. The contributions of diagram (9) are, respectively,
\begin{eqnarray}
&&\frac{2\alpha_{f}c^{3}}{\pi\epsilon_{0}e^{2}}\sum_{i<a,j<b}d_{ai}^{r}d_{jb}^{s}d_{jb}^{p}d_{ai}^{q}\int_{-\infty}^{+\infty}
\textrm{d}k\:k^{2}\textrm{Im}G _{rs}(\mathbf{r},k)\int_{-\infty}^{+\infty}\textrm{d}k'\:k'^{2}\textrm{Im}G _{pq}(\mathbf{r},k')\nonumber\\
&\times&\Bigl[\int_{-\infty}^{T}\textrm{d}t\int_{-\infty}^{t}\textrm{d}t'\int_{-\infty}^{t'}\textrm{d}t''\:e^{\eta(t+t'+t'')}
\Bigl(i\:e^{-i(T-t)(\omega+\omega_{bj})}e^{-i(t-t')(\omega+\omega')}e^{-i(t'-t'')(\omega+\omega_{ai})}
e^{-it''(\omega_{ai}+\omega_{bj})}\Bigr)\nonumber\\
&+&\int_{-\infty}^{T}\textrm{d}t\int_{-\infty}^{T}\textrm{d}t'\int_{-\infty}^{t'}\textrm{d}t''\:e^{\eta(t+t'+t'')}
\Bigl(i\:e^{i(\omega_{ai}+\omega_{bj})T}e^{-i(T-t)(\omega+\omega_{ai})}e^{-it(\omega_{ai}+\omega_{bj})}e^{i(T-t')(\omega+\omega')}
e^{i(t'-t'')(\omega+\omega_{bj}}
e^{it''(\omega_{ai}+\omega_{bj})}\Bigr)\Bigr]\nonumber\\
&=&\frac{-2\alpha_{f} c^{3}}{\pi\epsilon_{0}e^{2}}\textrm{Re}\sum_{i<a,j<b}d_{ai}^{r}d_{jb}^{s}d_{jb}^{p}d_{ai}^{q}\int_{-\infty}^{+\infty}
\textrm{d}k\int_{-\infty}^{+\infty}\textrm{d}k'\Bigl[\frac{k^{2}\textrm{Im}G _{rs}(\mathbf{r},k)\:k'^{2}\textrm{Im}G _{pq}(\mathbf{r},k')}
{(\omega+\omega'-\omega_{ai}-\omega_{bj}-2i\eta)(\omega-\omega_{bj}-i\eta)(\omega-\omega_{ai}-3i\eta)}\nonumber\\
&+&\frac{k^{2}\textrm{Im}G _{rs}(\mathbf{r},k)\:k'^{2}\textrm{Im}G _{pq}(\mathbf{r},k')}
{(\omega+\omega'-\omega_{ai}-\omega_{bj}+2i\eta)(\omega-\omega_{bj}-i\eta)(\omega-\omega_{ai}+i\eta)}\Bigr],\:\:\eta\rightarrow0^{+},\textrm{ to }\langle W_{A}/2\rangle_{\textrm{res}};
\end{eqnarray}
\begin{eqnarray}
&&\frac{2\alpha_{f} c^{3}}{\pi\epsilon_{0}e^{2}}\sum_{i<a,j<b}d_{ai}^{r}d_{jb}^{s}d_{jb}^{p}d_{ai}^{q}\int_{-\infty}^{+\infty}
\textrm{d}k\:k^{2}\textrm{Im}G _{rs}(\mathbf{r},k)\int_{-\infty}^{+\infty}\textrm{d}k'\:k'^{2}\textrm{Im}G _{pq}(\mathbf{r},k')\nonumber\\
&\times&\Bigl[\int_{-\infty}^{T}\textrm{d}t\int_{-\infty}^{t}\textrm{d}t'\int_{-\infty}^{T}\textrm{d}t''\:e^{\eta(t+t'+t'')}
\Bigl(-i\:e^{-i(T-t)(\omega+\omega')}e^{-i(t-t')(\omega+\omega_{ai})}e^{-it'(\omega_{bj}+\omega_{ai})}e^{i(T-t'')(\omega+\omega_{bj})}
e^{it''(\omega_{ai}+\omega_{bj})}\Bigr)\nonumber\\
&+&\int_{-\infty}^{T}\textrm{d}t\int_{-\infty}^{t}\textrm{d}t'\int_{-\infty}^{t'}\textrm{d}t''\:e^{\eta(t+t'+t'')}
\Bigl(-i\:e^{-i(\omega_{ai}+\omega_{bj})T}e^{i(T-t)(\omega+\omega_{ai})}e^{i(t-t')(\omega+\omega')}e^{i(t'-t'')(\omega+\omega_{bj})}
e^{it''(\omega_{ai}+\omega_{bj})}\Bigr)\Bigr]\nonumber\\
&=&\frac{-2\alpha_{f} c^{3}}{\pi\epsilon_{0}e^{2}}\textrm{Re}\sum_{i<a,j<b}d_{ai}^{r}d_{jb}^{s}d_{jb}^{p}d_{ai}^{q}\int_{-\infty}^{+\infty}
\textrm{d}k\int_{-\infty}^{+\infty}\textrm{d}k'\Bigl[\frac{k^{2}\textrm{Im}G _{rs}(\mathbf{r},k)\:k'^{2}\textrm{Im}G _{pq}(\mathbf{r},k')}
{(\omega+\omega'-\omega_{ai}-\omega_{bj}-2i\eta)(\omega-\omega_{ai}+i\eta)(\omega-\omega_{bj}-i\eta)}\nonumber\\
&+&\frac{k^{2}\textrm{Im}G _{rs}(\mathbf{r},k)\:k'^{2}\textrm{Im}G _{pq}(\mathbf{r},k')}
{(\omega+\omega'-\omega_{ai}-\omega_{bj}+2i\eta)(\omega-\omega_{ai}+i\eta)(\omega-\omega_{bj}+3i\eta)}\Bigr],\:\:\eta\rightarrow0^{+},
\textrm{ to }\langle W_{B}/2\rangle_{\textrm{res}};
\end{eqnarray}
and
\begin{eqnarray}
&&\frac{2\alpha_{f} c^{3}}{\pi\epsilon_{0}e^{2}}\sum_{i<a,j<b}d_{ai}^{r}d_{jb}^{s}d_{jb}^{p}d_{ai}^{q}\int_{-\infty}^{+\infty}
\textrm{d}k\:k^{2}\textrm{Im}G _{rs}(\mathbf{r},k)\int_{-\infty}^{+\infty}\textrm{d}k'\:k'^{2}\textrm{Im}G _{pq}(\mathbf{r},k')\nonumber\\
&\times&\int_{-\infty}^{T}\textrm{d}t\int_{-\infty}^{t}\textrm{d}t'\int_{-\infty}^{t'}\textrm{d}t''\:e^{\eta(t+t'+t'')}
\Bigl(i\:e^{-i(T-t)(\omega+\omega_{bj})}e^{-i(t-t')(\omega+\omega')}e^{-i(t'-t'')(\omega+\omega_{ai})}
e^{-it''(\omega_{ai}+\omega_{bj})}\Bigr)\\
&=&\frac{-2\alpha_{f} c^{3}}{\pi\epsilon_{0}e^{2}}\textrm{Re}\sum_{i<a,j<b}d_{ai}^{r}d_{jb}^{s}d_{jb}^{p}d_{ai}^{q}\int_{-\infty}^{+\infty}
\textrm{d}k\int_{-\infty}^{+\infty}\frac{\textrm{d}k'\:k^{2}\textrm{Im}G _{rs}(\mathbf{r},k)\:k'^{2}\textrm{Im}G _{pq}(\mathbf{r},k')}
{(\omega+\omega'-\omega_{ai}-\omega_{bj}-2i\eta)(\omega-\omega_{bj}-i\eta)(\omega-\omega_{ai}-3i\eta)},\:\eta\rightarrow0^{+},\nonumber
\end{eqnarray}
to $\delta\mathcal{E}_{\textrm{res}}$.

The corresponding contributions of diagram (10) are identical to those of diagram (9), but for the exchange of subindices $ai\leftrightarrow bj$ in all the expressions above.

Finally, for the case of two identical atoms excited, in the perturbative regime, we give the expressions of the contributions of diagram (4) of
Figs.\ref{fig1QED} and \ref{fig3QED}, with double poles, to $\langle W_{A}/2\rangle_{\textrm{res}}$ and $\delta\mathcal{E}_{\textrm{res}}$,
respectively, in the adiabatic approximation. These are,
\begin{eqnarray}
&&\frac{4\alpha_{f}c^{3}}{\pi\epsilon_{0}e^{2}}\textrm{Re}\sum_{i<a}d_{ai}^{r}d_{ai}^{s}d_{ai}^{p}d_{ai}^{q}\int_{-\infty}^{+\infty}
\textrm{d}k\:k^{2}\textrm{Im}G _{rs}(\mathbf{r},k)\int_{-\infty}^{+\infty}\textrm{d}k'\:k'^{2}\textrm{Im}G _{pq}(\mathbf{r},k')\nonumber\\
&\times&\int_{-\infty}^{T}\textrm{d}t\int_{-\infty}^{T}\textrm{d}t'\int_{-\infty}^{t'}\textrm{d}t''\:e^{\eta(t+t'+t'')}
\Bigl(-i\:e^{i(T-t)(\omega+\omega_{ai})}e^{2it\omega_{ai}}e^{-i(T-t')(\omega+\omega')}e^{-i(t'-t'')(\omega'+\omega_{ai})}
e^{-2it''\omega_{ai}}\Bigr)\label{l1}\\
&=&\frac{-4\alpha_{f} c^{3}}{\pi\epsilon_{0}e^{2}}\textrm{Re}\sum_{i<a}d_{ai}^{r}d_{ai}^{s}d_{ai}^{p}d_{ai}^{q}\int_{-\infty}^{+\infty}
\textrm{d}k\int_{-\infty}^{+\infty}\frac{\textrm{d}k'\:k^{2}\textrm{Im}G _{rs}(\mathbf{r},k)\:k'^{2}\textrm{Im}G _{pq}(\mathbf{r},k')}
{(\omega+\omega'-2\omega_{ai}-2i\eta)(\omega'-\omega_{ai}-i\eta)(\omega-\omega_{ai}+i\eta)},\:\eta\rightarrow0^{+},\nonumber
\end{eqnarray}
to $\langle W_{A}/2\rangle_{\textrm{res}}$; and
\begin{eqnarray}
&&\frac{4\alpha_{f}c^{3}}{\pi\epsilon_{0}e^{2}}\textrm{Re}\sum_{i<a}d_{ai}^{r}d_{ai}^{s}d_{ai}^{p}d_{ai}^{q}\int_{-\infty}^{+\infty}
\textrm{d}k\:k^{2}\textrm{Im}G _{rs}(\mathbf{r},k)\int_{-\infty}^{+\infty}\textrm{d}k'\:k'^{2}\textrm{Im}G _{pq}(\mathbf{r},k')\nonumber\\
&\times&\int_{-\infty}^{T}\textrm{d}t\int_{-\infty}^{t}\textrm{d}t'\int_{-\infty}^{t'}\textrm{d}t''\:e^{\eta(t+t'+t'')}
\Bigl(i\:e^{2iT\omega_{ai}}e^{i(T-t)(\omega+\omega_{ai})}e^{-i(t-t')(\omega+\omega')}e^{-i(t'-t'')(\omega'+\omega_{ai})}
e^{-2it''\omega_{ai}}\Bigr)\label{l2}\\
&=&\frac{-4\alpha_{f} c^{3}}{\pi\epsilon_{0}e^{2}}\textrm{Re}\sum_{i<a}d_{ai}^{r}d_{ai}^{s}d_{ai}^{p}d_{ai}^{q}\int_{-\infty}^{+\infty}
\textrm{d}k\int_{-\infty}^{+\infty}\frac{\textrm{d}k'\:k^{2}\textrm{Im}G _{rs}(\mathbf{r},k)\:k'^{2}\textrm{Im}G _{pq}(\mathbf{r},k')}
{(\omega+\omega'-2\omega_{ai}-2i\eta)(\omega'-\omega_{ai}-i\eta)(\omega-\omega_{ai}-3i\eta)},\:\eta\rightarrow0^{+},\nonumber
\end{eqnarray}
to $\delta\mathcal{E}_{\textrm{res}}$. In this case, the results of the frequency integrals in Eqs.(\ref{l1}) and (\ref{l2}) coincide.
\end{widetext}


\begin{thebibliography}{105}
\bibitem{Bermanbook} P.R. Berman, \emph{Cavity Quantum Electrodynamics}, Academic, New York (1994).
\bibitem{ReviewHarocheetal} J.M. Raimond,  M. Brune, and S. Haroche,  \emph{Rev. Mod. Phys.}\textbf{73}, 565 (2001).
\bibitem{expofHarocheBruneJMRaimon} F. Yamaguchi, P. Milman, M. Brune, J.M. Raimond, and S. Haroche,  \emph{Phys. Rev. A}\textbf{66}, 010302(R) (2002).
\bibitem{BuhmannScheel} S.Y. Buhmann and S. Scheel, \emph{Acta Phys. Slovaca} \textbf{58}, 675 (2008).
\bibitem{Barton} G. Barton, \emph{Proc. Roy. Soc. Lond. A}\textbf{367},  117 (1979); \textbf{410},  141 (1987).
\bibitem{Belov} A.A. Belov, Y.E. Lozovik, and V.L. Pokrovskii, \emph{Sov. Phys. JETP}\textbf{69},  312 (1989).
\bibitem{HindsSandoghdar} E.A. Hinds and V. Sandoghdar,  \emph{Phys. Rev. A}\textbf{43}, 398 (1991).
\bibitem{Jhe} W. Jhe,  \emph{Phys. Rev. A}\textbf{43}, 5795 (1991); \emph{Phys. Rev. A}\textbf{44}, 5932 (1991).
\bibitem{Lutken} C.A. L\"utken and F. Ravndal, \emph{Phys. Rev. A}\textbf{31},  2082 (1985).
\bibitem{MilonniKnight} P.W. Milonni and P.L. Knight, \emph{Opt. Commun.}\textbf{9},  119 (1973).
\bibitem{Nha} H. Nha and W. Jhe,  \emph{Phys. Rev. A}\textbf{54}, 3505 (1996).
\bibitem{Milonnibook}  P.W. Milonni,  \textit{The Quantum Vacuum}, Academic Press, San Diego (1994).
\bibitem{Haroche1} P. Nussenzveig, F. Bernardot, M. Brune, J. Hare, J.M. Raimond, S. Haroche and W. Gawlik, \emph{Phys. Rev. A}\textbf{48}, 3991 (1993); S. Haroche, M. Brune, and J.M. Raimond, \emph{Europhys. Lett.}\textbf{14}, 19 (1991); H. Walther, B.T.H. Varcoe, B.G. Englert, and T. Becke, \emph{Rep. Prog. Phys.}\textbf{69}, 1325 (2006).
\bibitem{Haroche} S.B. Zheng, and G.C. Guo, \emph{Phys. Rev. Lett.}\textbf{85}, 2392 (2000); S. Osnaghi, P. Bertet, A. Auffeves, P. Maioli, M. Brune, J.M. Raimond, and S. Haroche, \emph{Phys. Rev. Lett.}\textbf{87}, 037902 (2001).
\bibitem{Rydberg} L. B\'eguin, A. Vernier, R. Chicireanu, T. Lahaye and A. Browaeys, \emph{Phys. Rev. Lett.}  {\bf 110}, 263201 (2013).
\bibitem{WileySipe} J.M. Wylie  and J.E. Sipe,  \emph{Phys. Rev. A}\textbf{30}, 1185 (1984); \emph{Phys. Rev. A}\textbf{32}, 2030 (1985).
\bibitem{Miltonbook} K.A. Milton, \emph{The Casimir Effect: Physical Manifestations of Zero-point Energy}, World Sci., Singapore (2001).





\bibitem{McLone}  R.R. McLone and E.A. Power, \emph{Proc. R. Soc. A}\textbf{286},  573 (1965).
\bibitem{Craigbook} D.P. Craig  and T. Thirunamachandran, \emph{Molecular Quantum Electrodynamics}, Dover ed., New York (1998).
\bibitem{MePRA} M. Donaire,  \emph{Phys. Rev. A}\textbf{93}, 052706 (2016).
\bibitem{MylastPRA} M. Donaire,  \emph{Phys. Rev. A}\textbf{94}, 062701 (2016).
\bibitem{Pablo} P. Barcellona, R. Passante, L. Rizzuto, and S.Y. Buhmann, \emph{Phys. Rev. A}\textbf{94}, 012705 (2016).
\bibitem{MilonniPRA} P.W. Milonni and S.M.H. Rafsanjani, \emph{Phys. Rev. A}\textbf{92},  062711 (2015).
\bibitem{Berman} P.R. Berman, \emph{Phys. Rev. A}\textbf{91},  042127 (2015).
\bibitem{MePRL} M. Donaire, R. Gu\'erout and A. Lambrecht, \emph{Phys. Rev. Lett.}  {\bf 115}, 033201 (2015).







\end{thebibliography}
\end{document}